# A stochastic model for retinocollicular map development

Alexei A. Koulakov and Dmitry N. Tsigankov
*Cold Spring Harbor Laboratory, One Bungtown Road, Cold Spring Harbor, NY, 11724*

## Abstract

**Background**

We examine results of gain-of-function experiments on retinocollicular maps in knock-in mice [Brown et al. (2000) Cell 102:77]. In wild-type mice the temporal-nasal axis of retina is mapped to the rostral-caudal axis of superior colliculus (SC). The established map is single-valued, which implies that each point on retina maps to a single termination zone (TZ) in SC. In homozygous Isl2/EphA3 knock-in mice the map is double-valued, which implies that a single point on retina maps to two TZs in SC. This is a reflection of the fact that only about 50 percent of cells in retina express Isl2. In heterozygous Isl2/EphA3 knock-ins the map is intermediate between the homozygous and wild-type: it is single-valued in temporal and double-valued in the nasal parts of retina.

**Results**

We study the map formation using stochastic model based on Markov chains. In our model the map undergoes a series of reconstructions with probabilities dependent upon a set of chemical cues. Our model suggests that the map in heterozygotes is single-valued in the temporal region of retina due to reduced gradient of ephrin in the corresponding region of SC. The remaining map is double-valued since the gradient of ephrin is high there. We predict therefore that if gradient of ephrin is reduced by a genetic manipulation, the single-valued region of the map should occupy a larger portion of temporal retina, i.e. the point of transition between single- and doulble-valued maps should move to a more nasal position in Isl2-EphA3 heterozygotes. We also discuss the importance of inhomogeneous EphA gradient.

**Conclusion**

We present a theoretical model for retinocollicular map development, which can account for intriguing behaviors observed in gain-of-function experiments by Brown et al., including bifurcation in heterozygous Isl2/EphA3 knock-ins. The model is based on known chemical labels, axonal repulsion/competition, and stochasticity. Mapping in Isl2/EphB knock-ins is also discussed.

# Background

Topographic ordering is an important feature of the visual system, which is conserved among many visual areas [1]. Thus, the projection from retina to superior colliculus (SC) is established in a way, which retains neighbourhood relationships between neurons [2-4]. This implies that two axons of retinal ganglion cells (RGCs), which originate from neighbouring points on retina, terminate proximally in SC. It is assumed that this facilitates visual processing, which involves wiring local to the termination zone [5].

The mechanisms responsible for topographic ordering have been lately under thorough examination. Following the original suggestion by Sperry [6], it was shown that chemical labels play an essential role in formation of the map (reviewed in [3, 7]). For the projection from retina to SC the Eph family of receptor tyrosine kinases and their ligands ephrins-A were shown to be necessary for establishing correct topographic maps [7-10]. The coordinate system is encoded chemically in retina through graded expression of the Eph receptors by the RGCs. Thus, in mouse retina, two receptors of the family, EphA5 and A6, are expressed in the low nasal – high temporal gradient [11-14]. The recipient coordinate system in the SC is established through high caudal – low rostral gradient of ephrin-A2 and A5 ligands [15]. Since RGC axons expressing EphA receptors are repelled by high levels of ephrin-A ligands this dual system of gradients allows sorting of the projecting axons in the order of continuously increasing density of receptors, whereby contributing to the formation of topographic map [10, 15, 16] (Figure 1A). Thus, the dual gradient system is involved in formation of topographic representation along the nasal-temporal axis, albeit some additional tuning is provided by activity-dependent mechanisms [17-19].

In this study we address the results of gain-of-function experiments, in which the retinocollicular maps were modified by genetic manipulations [20]. RGCs of the wild-type mouse express the LIM homeobox gene Islet2 (Isl2) [21]. Retina of a single animal is composed of two types of cells with regard to their expression of Isl2 gene, Isl2+ and Isl2-, which are intermixes in roughly equal proportion throughout the RGC layer (Figure 1B). To test the mechanisms of the retinocollicular map formation Brown et al. [20] generated "knock-in" mice, in which the Isl2 and EphA3 genes are coexpressed. This implies that each Isl2+ RGC and its axons, in addition to EphA5 and A6, also expresses EphA3 gene, not found in the wild-type RGCs. The Isl2- cells remain EphA3-, as the wild-type cells. By doing so Brown et al. [20] increased the total level of EphA receptors in a given fraction of retinal cells. Since the overall level of EphAs is increased in Isl2+/EphA3+ cells, axons of two neighboring cells, knock-in and wild-type, should terminate in quite different places in SC (Figure 1B). The knock-in cell, interacting more strongly with the repellent should terminate at the position of decreased density of ephrins, i.e. more rostrally with respect to the wild-type cells. The neighborhood relationships between axons should be lost, the new map should lose its continuous nature, and it should split into two maps: one for wild-type RGCs, one for knock-in cells. This prediction was confirmed by experiments of Brown et al. (Figure 2).

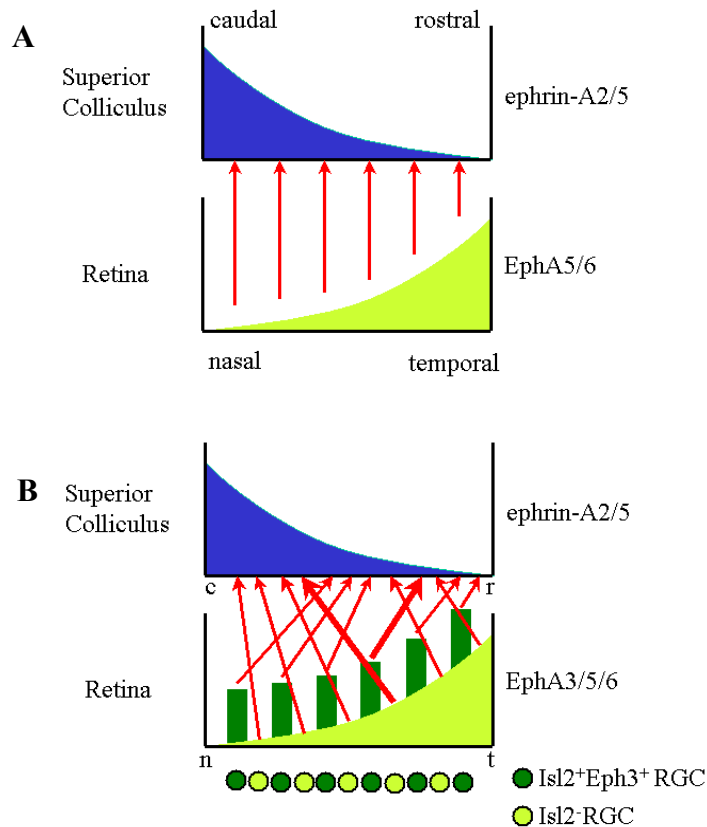

**Figure 1 - Chemical labelling system in retinocollicular map in mice**

A. Formation of the map in the wild-type mouse. Retinal ganglion cells (RGC) express EphA5/6 receptors in temporal > nasal gradient (bottom), whereas the cells in SC express the ephrin-A ligands in caudal > rostral gradient (top). Since axons of Eph+ RGC (red arrows) are repelled by ephrins this distribution of chemical markers leads to establishing of ordered topographic map in which nasal/temporal retina projects to caudal/rostral SC. This is because RGC axons expressing highest levels of Eph receptors (temporal) experience the largest repulsion and are expelled to the rostral part of SC, where such repulsion is minimal. Axons of nasal RGC are more indifferent to the action of ligands and occupy more caudal positions. Such system allows positioning of RGC axons in the order of increasing expression level of EphA receptors.

B. Map in the mutant mouse from Ref. [20]. The expression level of EphA receptors was artificially increased in every second cell by genetic manipulations (dark gray). This is done by co expressing EphA3, which is absent in the wild-type RGCs (see **A**), with another gene, Isl2, which is expressed roughly in 50% of RGCs. Since ephrin ligands bind and activate all receptors from EphA family, albeit with different affinity, this results in anomalous projection to SC, based roughly on the total levels of EphA in each axon. Similarly to **A** this leads to sorting of axons in the order of increasing density of EphAs (red arrows). Note that two RGC neighboring in retina become separated in SC (bold arrows). This aberration in the topographic map leads to two termination zones (TZs) in SC for two neighboring cells in retina, rather than a single zone in wild-type [20].

In addition to the observation of the overall map doubling in homozygous knock-ins (Figure 2C), Brown et al. discovered a curious behavior of the map in heterozygous animals. In these animals the exogenous levels of EphA3 were reduced roughly by a factor of two with respect to the homozygous knock-ins (Figure 2B). In terms of the expression density of EphA3 these animals stand between the wild-type and knock-in animals. Accordingly, the structure of the map resembles a hybrid of the wild-type and homozygous maps. The more rostral part of the map is single-valued, similarly to the wild-type, whereas about 60% of the caudal-most part is double-valued, like in the homozygous animals. This observation suggests that the map bifurcates somewhere between double- and single-valued regions. Although overall doubling of the map in homozygotes is easy to understand, any true model for the retinocollicular map formation should be able to account for the bifurcating behavior of map in heterozygotes. Therefore, experiments in heterozygotes represent a powerful tool to falsify various theoretical models.

Ref. [20] suggests that the bifurcating behavior of the map is consistent with the importance of relative rather than absolute values of the expression levels. Indeed, the relative difference of exogenous EphA3 to endogenous EphA5/6 is maximal in nasal retina (caudal SC), where the doubled map is observed (Figure 2B). In the temporal retina (rostral SC) the EphA3 to EphA5/6 ratio is not so large, which may account for the fact that the map is single-valued there. Thus a model for the topographic map from retina to SC should rely on the relative but not absolute levels of EphA signaling.

The point, which we make in this study, is that more experimental tests are needed to justify the suggestion about relative expression levels. To make our point clear we present a model for the retinocollicular map formation, which is based upon differences in the <u>absolute</u> values of Eph/ephrin expression levels, rather than relative differences. Our model manages to reproduce all the essential features of experiments described in Ref. [20], including bifurcation of the map in heterozygotes. This model was previously reported in [22]. In the model presented here the map is single-valued in rostral part of heterozygous SC due to inhomogeneous gradient of ephrin ligand and its receptor, rather that reduced relative difference of EphA receptors. Below we suggest experimental tests, which may distinguish these two classes of models.

To quickly test various hypotheses we developed a model for retinocollicular map formation employing stochastic Markov chain process (explained below). Our model is based upon three principles: chemoaffinity, axonal competition, and stochasticity. The implementation of the model used here is available in [23].

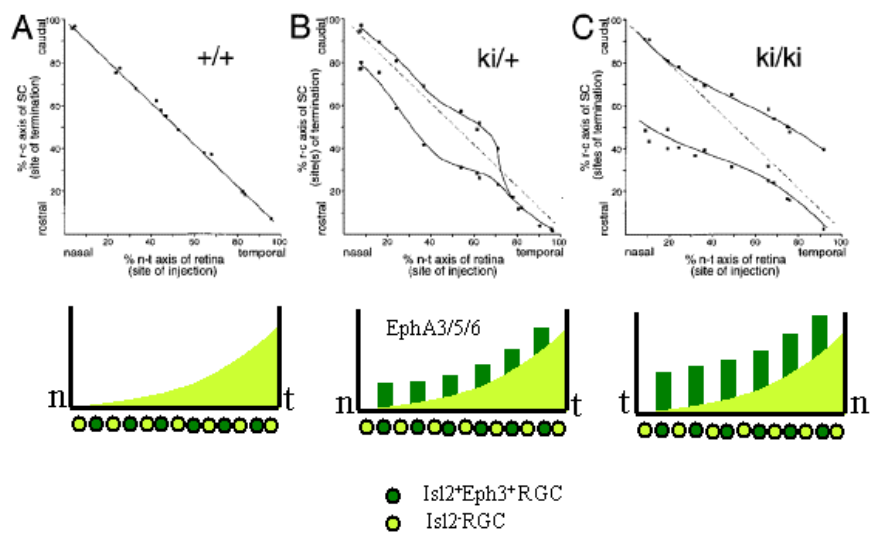

**Figure 2 - Maps in Isl2+/Eph3+ mice**

The top row is reproduced from Ref. [20] (Figure 5). The bottom row illustrates the corresponding distribution of EphAs.

# Results

**Markov chain model**

Let us first describe the 1D version of the model. We consider a linear chain of 100 RGC, each expressing individual level of EphA receptors given by function $RA(i)$, where $i = 1...100$ is the RGC index, which also determines a discrete position of the cell in the retina. We have verified that results presented below do not depend on the number of cells, as long as this number is large enough. Each RGC is attached by an axon to one and only terminal cell in SC, which has an expression level of ligand given by $LA(k)$, where $k = 1...100$ is the index in SC, also describing the terminal position. The receptor density $RA$ is an overall increasing function of its index $i$, while the ligand density $LA$ is decreasing, when going from $k = 1$ (caudal) to $k = 100$ (rostral) positions (Figure 3). This determines the layout of chemical "tags" used to set up map's "topography". An additional feature is that no two cells can project to the same spot in SC, which is meant to mimic axonal repulsion/competition for positive factors in SC, described in detail by Refs. [10].

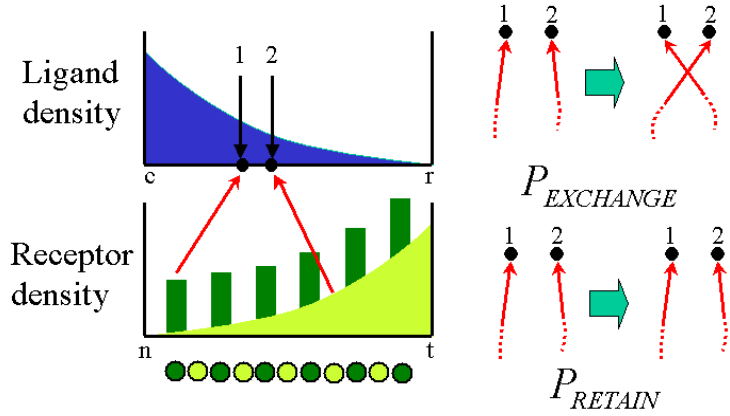

**Figure 3 - Description of the 1D model**

We then modify the map probabilistically, using the following rule. We consider two axons projecting to the neighboring points in SC (1 and 2 in Figure 3). We attempt to exchange these axons in SC with probability

$$P_{EXCHANGE} = \frac{1}{2} + \alpha [RA(1) - RA(2)][LA(1) - LA(2)] \qquad (1)$$

Here $\alpha > 0$ is the only parameter in our model. The probability of the axons to stay unchanged $P_{RETAIN}$ is determined from $P_{EXCHANGE} + P_{RETAIN} = 1$ and is therefore given by

$$P_{RETAIN} = \frac{1}{2} - \alpha [RA(1) - RA(2)][LA(1) - LA(2)]. \qquad (2)$$

Since the only difference between these probabilities is the sign in front of $\alpha$, it is important to understand the nature of this sign.

Assume that the product of gradients in Eq. (1) is negative, i.e. the gradients run in the opposite directions, which corresponds to the correct order of axonal terminals in SC. Then $P_{EXCHANGE} < 1/2$ and $P_{EXCHANGE} < P_{RETAIN}$, i.e. the probability or retaining the current ordering of the axonal pair is larger than changing it. This is consistent with the chemorepellent interactions of receptors and ligands. In the opposite case of the wrong order, i.e. when the product of gradients in Eq. (1) is positive and gradients run in the same directions, $P_{EXCHANGE} > P_{RETAIN}$ by the same reasoning. The described process will tend to exchange the order of gradients and, therefore establish the right order of topographic projections. By using probabilities described by Eqs. (1) and (2) we incorporate the chemoaffinity pronciple into our stochastic model. The step is then repeated for another nearest neighbor couple, chosen randomly, and so on, until a stationary distribution of projections is reached.

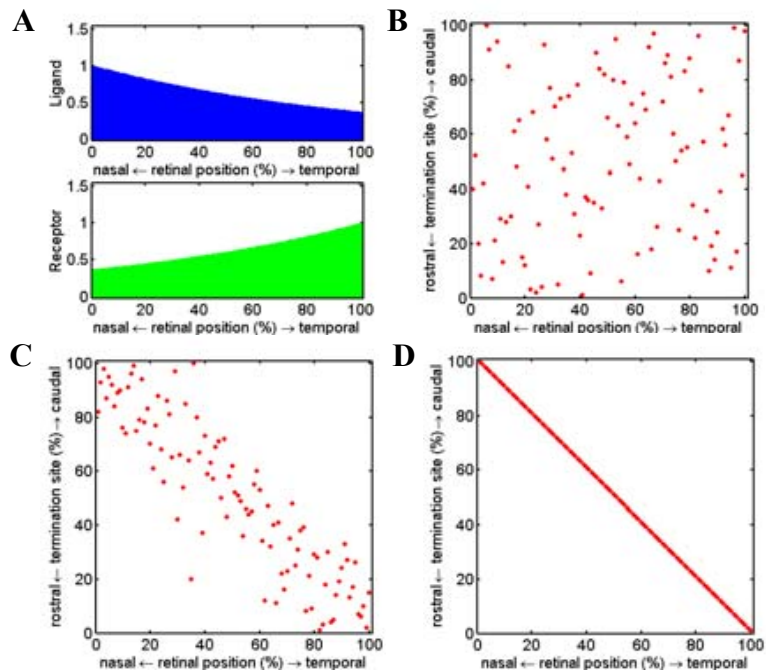

**Figure 4 - Typical solutions of 1D model**

**A**. Distribution of ligand (top) and receptor (bottom). **B**. Solution of the model for $\alpha = 0$. Red dots represent terminal positions of individual axons, originating at various points in retina. If the case $\alpha = 0$ the map is completely random, since all the chemical cues are multiplied by zero in Eq. (1), and, therefore, cannot contribute to the solution. **D.** $\alpha$ is very large. Solution represents perfect ordering of axons in SC in the order of increasing density of receptor. This is because the chemical cues are extremely strong in this case, much stronger than noise. **C.** $\alpha = 30$. At the intermediate value of $\alpha$ solution is a compromise between chemical signal and noise.

How unique is the choice of probabilities (1) and (2)? Our investigations show that (1) and (2) describe a very broad class of models, which tend to align gradients of receptor and attractor levels in the opposite directions. One can certainly imagine a more complicated function of densities than (1) (see also Discussion), but this more sophisticated function could be reduced to (1), which becomes the first order of the Taylor expansion. The latter expansion is valid if noise in the system is small enough, so that the size of the projection spot is much smaller than dimensions of retina. This technical point will be discussed in [24].

The process defined by (1) belongs to the class of Markov chain processes, invented by a Russian mathematician Andrei Markov. This class is general enough to describe many processes in biology, chemistry, and physics, such as ion channel kinetics, probabilistic synaptic models [25], processes of Brownian motion [26], etc. Markov chain is defined as a process in which probability of transition from one state to another is determined only by the present state of the system. In other words, this is a process in which memory of distant past is irrelevant. The most important for our purposes is the theorem about Markov processes of this sort [27] called ergodic theorem. It states that the process will eventually converge to some final stationary distribution of probabilities [26]. This is in contrast to other processes, in which distributions could go through cycles as a function of time or collapse to a single solution. Such property of ergodicity is obviously very helpful when considering the processes of development. Below we calculate the final distributions of probabilities as well as the intermediate states of the system in development.

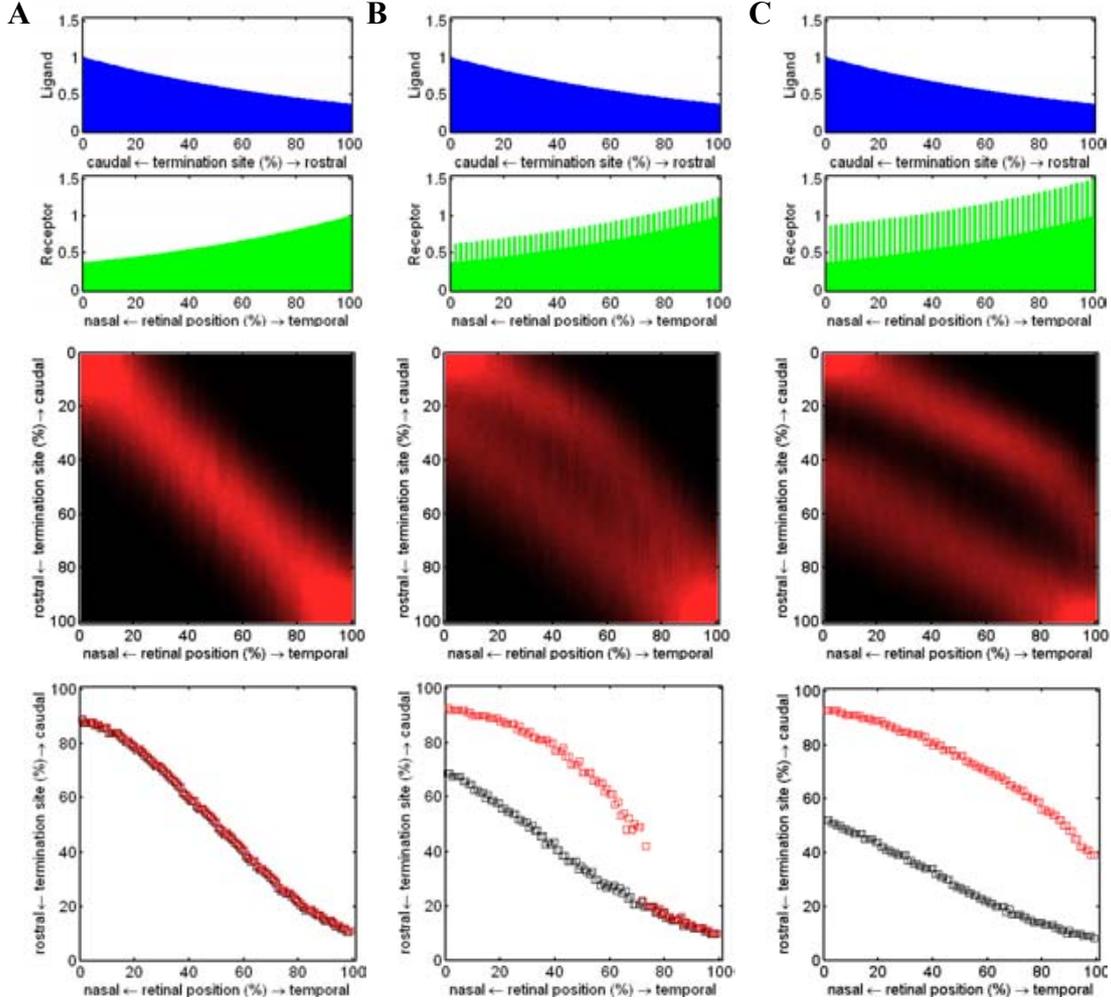

**Figure 5 - 1D maps in wild-type and knock-in animals.**
The top row represents distributions of chemical labels. The middle row shows the probability distributions, which are obtained from scattered plots similar to 4C by averaging over many trials. The brightness of color at each point in the middle image represents probability that an axon originating from the point's abscissa projects to the point's ordinate. The bottom row shows positions of maxima of probability density distributions above. The red markers correspond to wild-type cells; black markers determine maxima of distributions for Isl2+/EphA3+ RGCs. **A.** column of results for wild-type conditions. **B.** Column corresponding to heterozygous Isl2/Eph3 knock-in conditions. The bifurcation of the probability distribution is similar to results of Ref. [20] (cf. Figure 5 or Figure 2 above). **C.** Results for homozygous knock-in conditions.

To understand these results better we suggest to consider cases in which solution is obvious and can be obtained exactly. The model described by (1) can be solved exactly for at least two limiting cases: $\alpha = 0$ and $\alpha$ is very large. In the former case the information about chemical labels cannot affect the solution, since it is multiplied by 0 in Eq. (1). Hence, the map is completely <u>random</u> in this case (Figure 4B). In the latter case the molecular cues are very strong. They eventually produce solution in which the axons are perfectly <u>sorted</u> in SC in the order of increasing density of receptor (Figure 4D). An intermediate situation with certain finite value of parameter $\alpha$ is described by a compromise between noise and chemical cues, with former randomizing the map on the finer scale, while the latter inducing the overall correct ordering (Figure 4C). We conclude that mean position of projections is controlled by the chemical signal, while the <u>spread</u> of projections or the size of TZ is determined by noise (Figure 4C).

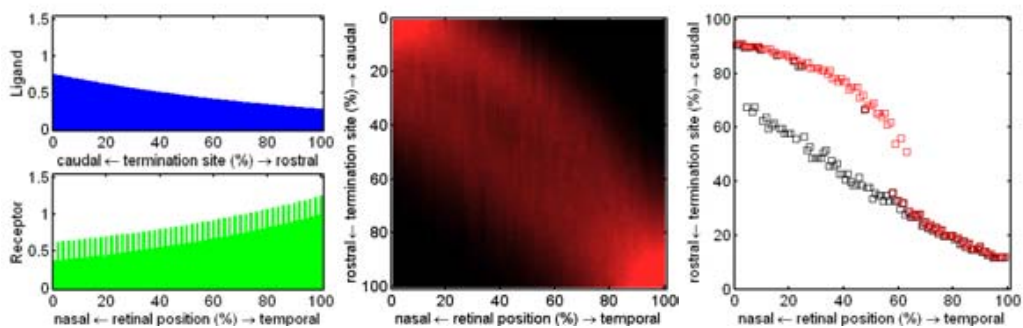

**Figure 6 - Results for heterozygous conditions and reduced by 25% density of ligand**
The bifurcation point is shifted caudally, compared to Figure 5B, as predicted by increased impact of noise. Notice that the bifurcation transition is discontinuous, as in Figure 5.

The notion of interactions between signal and noise is important in understanding maps in <u>heterozygotes</u>. Indeed, maps in both wild-type mice and homozygotes can be trivially understood on the basis of axonal sorting in SC in the order of monotonously increasing levels of EphAs. This is clear from Figures 1A and B. The impact of noise in those two cases is simply to produce some scattering of axons and smear termination zones, as explained in the previous paragraph. However, in case of heterozygotes, noise leads to qualitative changes in the map (Figure 5B). In this case the distance between two branches of the map, wild-type and EphA3+, is intermediate between the wild-type and homozygotes. Indeed, the distance between maps in homozygotes is about 40% (Figure 5C); in heterozygotes it is about twice as small. The chemical signal, in the form of distance between two maps, is therefore reduced in heterozygotes by a factor of two with respect to homozygotes. But the strength of noise is preserved, at least, in our model, since $\alpha = 30$ is all three cases in Figure 5. Therefore smearing of the map remains the same. Thus, two maps acquire potential to blend, as we see in Figure 5C.

Why does the blending of two maps first occur in rostral SC? Our analysis shows that two factors contribute to this phenomenon. The first factor is reduced gradient of ligand in rostral SC. The second factor is inhomogeneous endogenous EphA5/6 density. Both these factors are discussed below in some detail.

Let us first discuss the impact of ligand. The degree of noise is determined not only by parameter $\alpha$ but also by the gradient of ligand, since the smaller the gradient of ligand, the more disordered is the map [10]. But the gradient suffers significant reduction in rostral SC, at least in our model. Therefore, noise is stronger in rostral SC, and the size of the TZ is the largest there. To see this examine the TZs obtained in a 2D simulation (described below) in Figure 7. The lowest central picture shows two TZs for homozygous animal. The more rostral TZ has the largest extension in the rostral-caudal direction. But larger rostral TZs can blend sooner that smaller caudal ones. Therefore for given equal distance between two maps blending is more likely to occur in rostral rather than in caudal SC. We conclude that in our model the single-valued region in rostral SC is in part due to smearing of the two maps by noise.

To illustrate once again that the bifurcation is controlled by the inhomogeneous gradient of ligand in SC we reduce the density of ligand in SC uniformly by 25% (Figure 6). Since reduced gradient entails increased impact of noise and enlarged TZs, we expect the single-valued part of the map to expand. This prediction is confirmed by the numerical experiment (Figure 6). Notice that the only difference between Figure 5B and 6 is the reduced density of ligand. The point of transition is located more caudally in Figure 6. Hence, one can affect the transition in a predictable fashion by changing the density of ligand only.

Let us now briefly discuss the impact of EphA distribution on bifurcation. Assume for a moment that the impact of noise is very small. Then receptor is perfectly sorted in colliculus. This situation is described approximately by Figure 5C (bottom). One can observe that the distance between two maps is smaller in rostral than in caudal part. Thus, chemical signal itself is reduced there. This occurs because the gradient of ligand is inhomogeneous in retina. We further discuss why this happens in Discussion.

**Results for 2D model**

We simulated 2D development using the hypothesis that another pair of chemical tags, EphB family of receptors and their ligands, ephrins-B, are responsible for establishing topographic projection from dorsal-ventral (DV) axis on retina to lateral-medial axis in SC (see [9] for review). EphB2/3/4 are expressed in high-ventral-to-low-dorsal gradient by RGCs [28-30], while ephrins-B are expressed in high-medial-to-low-lateral gradient in tectum/SC [30]. Since dorsal/ventral axons project to lateral/medial SC this implies attractive interactions between EphB+ axons and ephrin-B rich environment [31] (see, however [32]). In our model the attractive interactions are modeled by the following exchange probability of two axonal terminals in the DV direction:

$$P_{EXCHANGE} = \frac{1}{2} - \beta [RB(1) - RB(2)][LB(1) - LB(2)] \qquad (3)$$

Here $RB(1)$, $RB(2)$, $LB(1)$, and $LB(2)$ are EphB receptor and ephrin-B ligand densities at neighboring points 1 and 2 in SC. This probability is similar to Eq. (1). Notice a sign change compared to Eq. (1), which insures that $P_{EXCHANGE} > P_{RETAIN}$ if the order of gradients is wrong, i.e. if the gradients of receptor and ligand are antiparallel. By choosing this sign we therefore ensure attraction between axons and ligands.

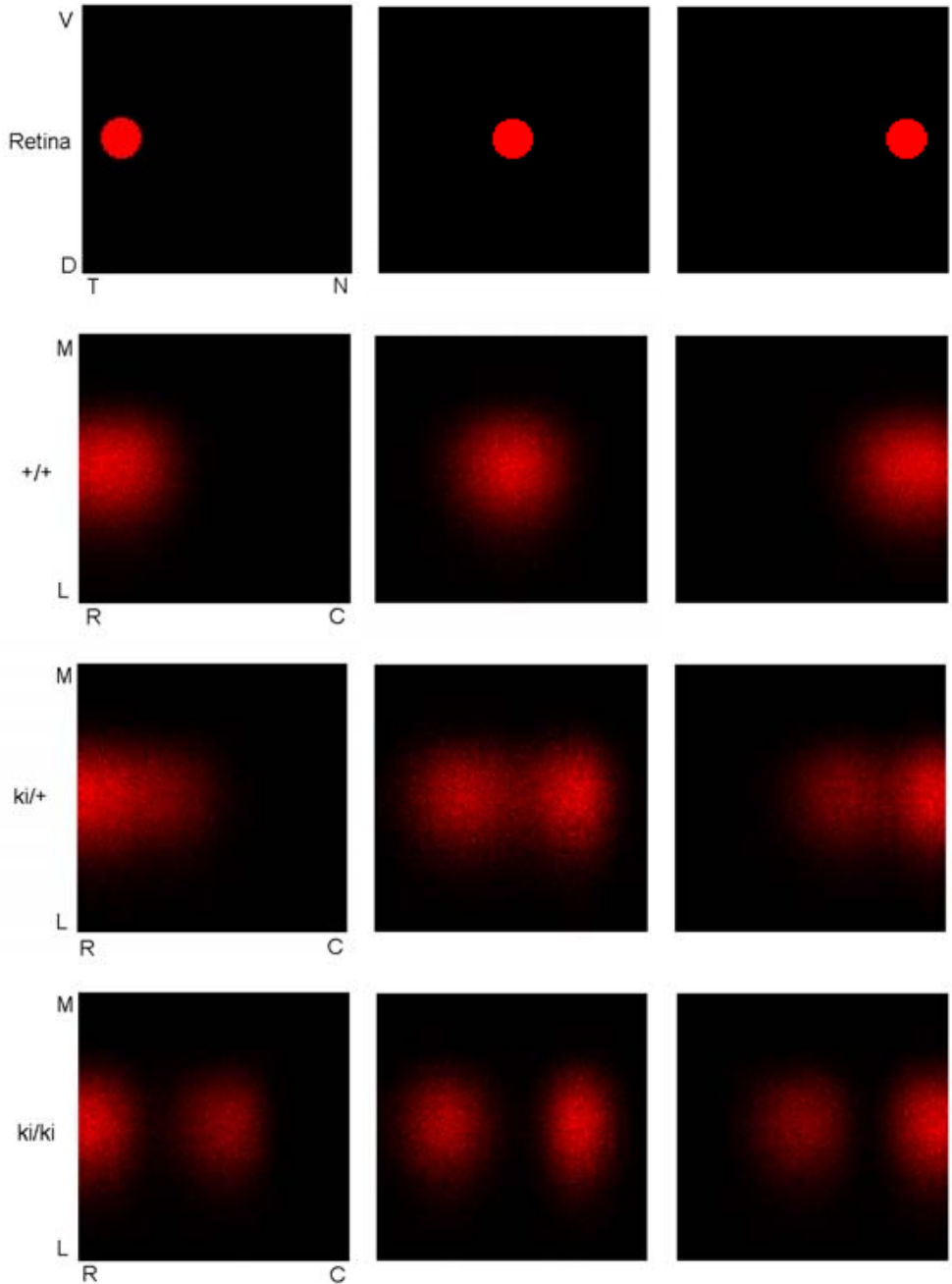

**Figure 7 - Numerical simulation of labelling in the 2D model.**
The top row shows anterograde "labeling spots" in retina. The following three rows display corresponding distribution of label in SC. The size of both retinal and collicular arrays is 100 by 100 cells. The three rows show results for wild-type, heterozygous knock-ins, and homozygous knock-ins, as marked on the left. Notice the doubling transition, when going from temporal to nasal injection in heterozygotes. This Figure is to be compared to Figure 4 from Ref. [20]. $\alpha = \beta = 30$. The color map is shifted in each image for visibility.
Abbreviations: D, dorsal; V, ventral; N, nasal; T, temporal; C, caudal; R, rostral; L, lateral; M, medial.

The details of our simulations are described in Methods. Our model allows not only exploration of two-dimensional maps (Figure 7) but also observing and modeling temporal development (Figure 8). Videos with detailed evolution of the map are available in [23].

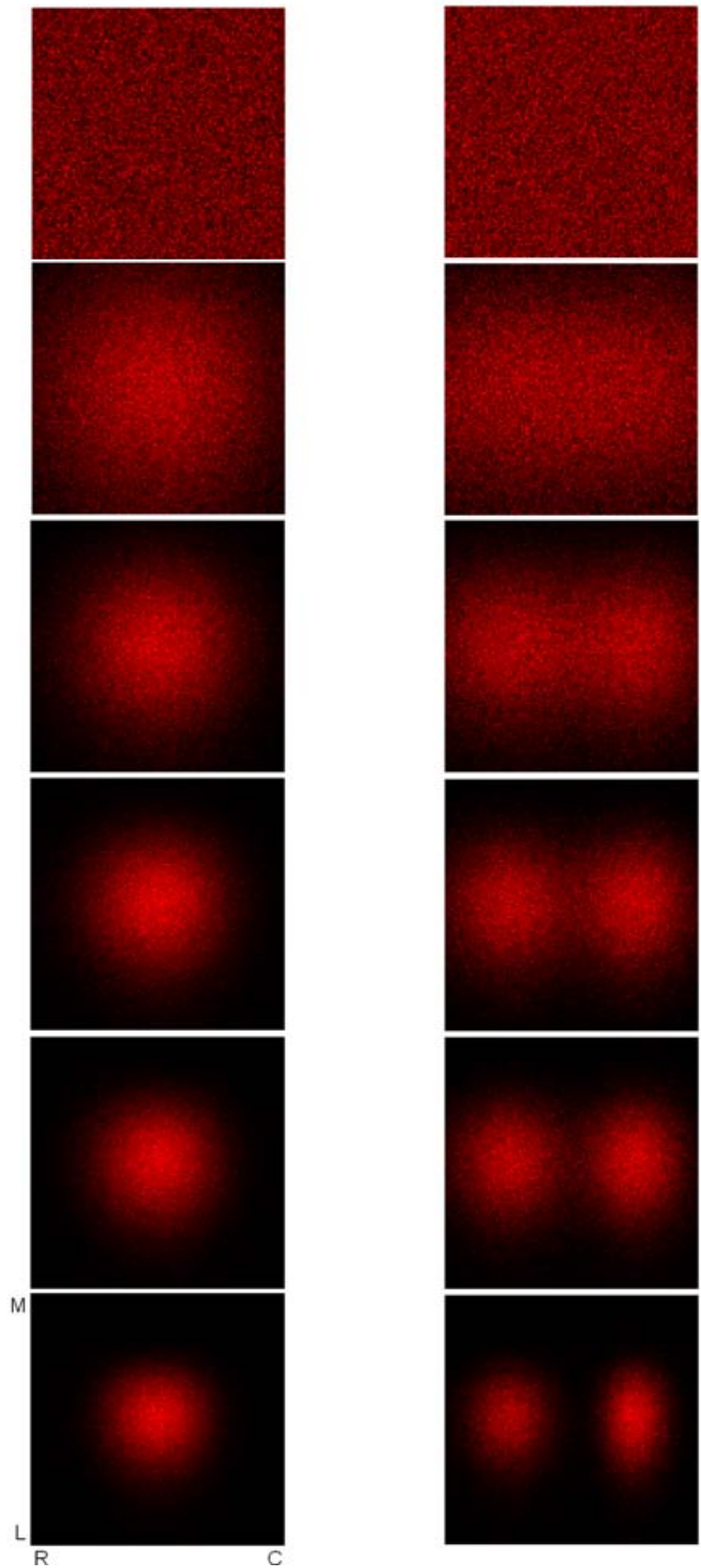

**Figure 8 - Map's refinement**

Wild-type (left) and ki/ki (right) map development for axons in the central retina. The retinal injection site is the same as in Figure 7, top row and central column. The temporal evolution of the map is shown for t = (0, 8, 16, 24, 32, and 100) × 10000 iterations. The orientation of images is the same as in Figure 7. The color map is rescaled in each image for visibility.

# Discussion

**Why does the map in heterozygotes bifurcate?**

In our model the map is formed through an interaction of three factors: repulsion/competition between axons for space (discussed in [10]), Eph/ephrin chemorepulsion/attraction, and noise. It is the latter that mixes two maps together in the rostral SC (Figures 5B, 6, 7). To understand how this behavior emerges in our model consider the main equation of our model (1) (Figure 9A). The part of the equation, which is proportional to $\alpha$, carries information about chemical signal, while ½ is responsible for noise. The chemical signal is also proportional to the gradient of ephrin ligand in SC, $LA(1) - LA(2)$. Here 1 and 2 are two neighboring receptacles. But the gradient of ligand is smaller in rostral than in caudal SC (Figure 9B). Hence, signal is smaller in rostral part and the impact of noise is the largest there. Noise mixes two maps (wild-type and knock-in) and produces a single-valued map in rostral SC. The hypothesis that impact of noise is stronger in rostral SC is consistent with ephrin-A knock-out experiments [10], since the effects of reduced density of ligand first occur in rostral SC.

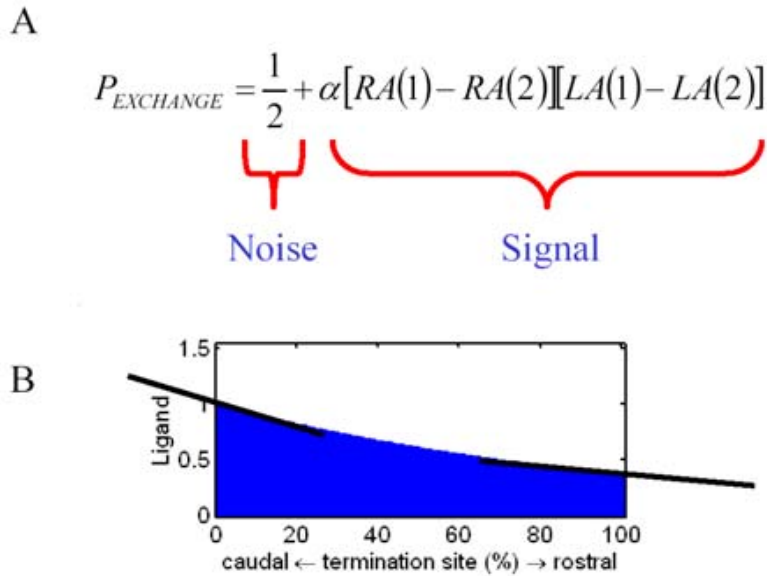

**Figure 9 - Interplay between signal and noise.**

To illustrate that the distribution of ligand controls doubling in our model, we "knock-out" some of the ligand in "SC" in our numerical simulation by reducing the ligand density by 25% (Figure 6). We observe a caudal displacement of the bifurcation point. Such manipulation could be done if Isl2/Eph3 knock-ins [20] are crossed with ephrin-A knock-outs [10].

The second factor contributing to bifurcation in heterozygotes is inhomogeneity of EphA gradient in retina. Consider the case of no noise. Mapping is obtained by sorting axonal terminals in the order of increasing density of EphA (Figure 10). Separation between two maps is smallest in the rostral part (Figure 10B). This is because of inhomogeneous gradient of receptor in "retinal" cells (Figure 10A). Therefore, even if noise were the same in all parts of the map, rostral part has the smallest signal in terms of separation between two maps, and the largest potential to be blended by noise. Of course all these claims apply only to this model.

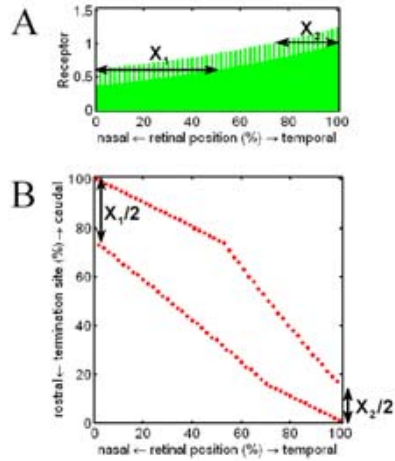

**Figure 10 - Mapping for heterozygotes in case of no noise.**
The separation between wild-type and EphA3 knock-in axonal terminals is larger in caudal than in rostral SC, due to inhomogeneity in EphA profile.

We conclude that two factors, increased noise and reduced signal, cooperate in rostral SC. This leads to the formation of single-valued map. In caudal part both noise is reduced and distance between maps is larger. Hence, the map is double-valued. Is it possible to distinguish these two factors? To do so we performed a numerical experiment on the Isl2/EphB "knock-in" conditions. This may have relevance to mapping in DV direction. The results are shown in Figure 11.

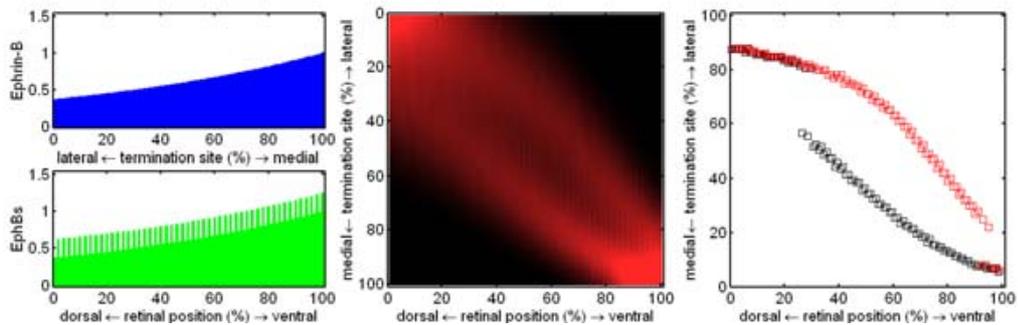

**Figure 11 - Mapping in Isl2/EphB knock-ins.**
Two bifurcations are observed, one in medial, another in lateral SC.

Two bifurcations observed in Figure 11 confirm the hypothesis about two factors. The ventral bifurcation is associated with receptor, since separation between two maps in perfectly ordered conditions is the smallest in medial SC. The second bifurcation, dorsal, occurs due to noise, since noise is maximal where the gradient of ligand is the smallest, i.e. in lateral SC. Thus, we suggest that experiments on Isl2/EphB knock-ins should make clear if inhomogeneity in receptor density or noise is more important.

**Absolute versus relative**

Ref. [20] undoubfully demonstrates that retinocollicular mapping is based on relative levels of EphA/ephrin-A expression in the broad meaning of this term. Indeed, the absolute value of EphA density does not determine where an axon terminates in colliculus. In fact, wild-type axons terminate more caudally in the presence of Isl2+/EphA3+ axons. Thus, an important factor is the presence of other axons, relative to which given axon establishes its termination point. This idea is also evident from retinal and collicular/tectal ablation experiments in rodents [33, 34] and other species [2].

Can we take this idea to the next level and say that relative differences between neighboring retinal cells represent the chemical signal? In this study we present a model, which takes <u>absolute</u> values of the chemical label as input, as seen from Eq. (1). In our model adding a constant value to all densities does not change the result, since (1) depends only on differences in expression levels. But this manipulation decreases the relative differences in the expression of EphAs between knock-in and wild-type cells. Hence, our model is not based on relative differences between receptor densities. Yet, we demonstrate that it can account for experimental results in detail. Thus, we suggest that existing experimental evidence is not sufficient to distinguish relative and absolute labeling in the narrow sense, suggested by Ref. [20], if we are to interpret these findings conservatively.

Of course, our model also accounts for the caudal displacement of wild-type TZs, thus resulting in a relative labeling system in the broad sense. In the first approximation, this model performs a sorting procedure, understood mathematically, of the fibers based on the expression levels of EphA. Our procedure uses differences in absolute values of EphA densities rather than relative differences. More experiments are needed to distinguish these two "relativity principles".

Relative labeling in the narrow sense can be incorporated in our model, if coefficient $\alpha$ is a function of label densities. Thus, the condition $\alpha \propto 1/(RA \cdot LA)$ ensures the Weber's law for axonal "perceptual thresholds", since chemical signal is proportional to the relative differences. The exact form of coefficient $\alpha$ or if the nomenclature based on this coefficient can account for all phenomenology is impossible to establish at the moment.

**On the biological realism**

When dealing with numerical simulations one always faces the question of the degree of realism with which to model the data. Does one have to model behaviors of individual atoms, or description on the level of axons is sufficient? In this work we choose the level of description on the basis of what we know. We realize that our model does not capture many exciting behaviors, but we argue that the mechanisms involved are unclear at the moment to be incorporated into a more detailed model. Our approach also fulfils its original goal, which is to reproduce the results of experiments [20] and to generate experimentally testable predictions, thus satisfying the requirement of parsimony.

Model presented here does not describe difference between development along TN and DV axes. The former mapping is controlled by original axonal overshoot along the RC direction in SC, with subsequent elimination of topographically inappropriate projections [3, 9]. In contrast, axons from the same DV retinal position enter SC in a broad distribution along ML axis. Topographically precise termination is provided by producing additional interstitial branches in the ML direction [31, 32, 35, 36]. These findings cannot be reproduced by our model, since no distinction is made between the original RGC axon and its branches. Instead, our model deals with terminal points of interstitial branches produced by RGC axons. It cannot address the mechanism by which this point is connected to a RGC.

Above we presented some results for mapping in 2D. The exact mechanism of mapping in 2D is unclear. We introduce this new component into our consideration for illustrative purposes only. Some curious observations can be made however. It was suggested earlier, that ephrins-B must have a bifunctional action on interstitial branches, both as a repellent and attractant, to implement topographic mapping in ML direction (see Ref. [32] for elaborated discussion of this point). Interstitial branches, which originate laterally from TZ must be attracted up the gradient of ephrins-B, while branches originating medially, must be repelled toward TZ by high density of ligand. Although we do not treat interstitial branches in our model, we can observe the location of their terminal points as a function of time (Figure 8). Indeed, the termination points located medially from the TZ are retracted and inserted more laterally in their drift toward TZ, thus moving <u>opposite</u> to the gradient of ephrin-B (Figure 8). In our model this behavior is mediated by axonal competition for space, since each axon can occupy only one terminal in SC. Indeed, axons with higher levels of EphB experience higher attraction to the ligand, expelling other axons to the region with lower density of ligand, effectively inducing chemorepulsion. This phenomenon is similar to behavior of passengers in a subway during rush hour. Although many passengers attempt to enter a newly arrived train, only limited number of them can, due to excluded volume interactions. The less motivated individuals left on the platform may seem to be repelled by the train, which is, of course, purely spurious. Thus, excluded volume interactions may generate effective repulsion by an attractive agent. This is what is observed on Figure 8 in our model.

## Conclusions

We present a model for retinocollicular map development, which can account intriguing behaviors observed in gain-of-function experiments by Brown et al. [20], including bifurcation in heterozygous Isl2/EphA3 knock-ins. The model is based on chemoaffinity, axonal repulsion/competition, and stochasticity. We discuss mapping in ephrin-A-/Isl2+/EphA3+ knock-out/ins and Isl2/EphB knock-ins.

## Methods

**1D model**

To find a stationary distribution of the RGC's axons in the SC, we use the following computational procedure. We consider a linear chain of 100 RCG that are connected to one and only terminal cell in SC each. The receptor and ligand expression level profiles used in the computations for wild-type, heterozygote and homozygote are shown at Figure 5A-C. We start with the random map where the position of every axon in SC does not depend on the level of its receptor expression. Then we perform stochastic reconstructions through an exchange of the positions of the neighboring axons in SC. Namely, at each step we randomly choose one pair of axons out of 99 neighboring pairs and switch their positions with the probability given by Eq. (1). In both cases, whether the positions of the axons are exchanged or they retain at their old locations we proceed to the next step when we choose a new pair of neighboring axons. We repeat the process until a stationary distribution of the probabilities for the positions of the RGC's axons in SC is reached.

The typical stationary solution for one realization is shown at Figure 4. Here the number of iterations is $10^6$. The only parameter of the theory is taken $\alpha=30$ throughout the paper. It is chosen to fit the experimental data from Ref. [20]. The probability distribution and the position of the maximums shown at Figure 5 and 6 are obtained by temporal averaging over $5 \times 10^4$ realizations of stationary solution separated in time by $10^3$ iterations.

**2D model**

Here we describe our 2D model in more detail. We consider an array of 100 by 100 RGC, which are connected to 100 by 100 different points in colliculus. Each RGC is characterized by two levels of expression for two receptors, EphAs and EphBs, described in the text. The concentration profiles are taken to be the same for EphA and EphB receptors in the wild-type species. In the homozygote and heterozygote cases the concentration of EphA is taken as shown at Figure 5, while the concentration of EphB is unchanged. RGCs do not express ligand in our model. The collicular receptacles are described by two ligand concentrations with the same profiles as shown at Figure 5 but with different gradient directions discussed in the text.

The process of development is modeled as follows. We randomly choose a pair of axons in SC separated either in RC or in ML direction. We exchange their positions with the probability given by Eq. (1) or Eq. (3) respectively. We then repeat the process until a stationary distribution of probabilities is reached in the same manner as for 1D case. Note, that this time a chosen pair of axons, say in RC direction, may not be a neighboring pair, but consist of two axons separated by any distance in SC. This procedure dramatically decreases the convergence time to the stationary distribution, which is the same as in the case when we choose the neighboring axons only. The discussion of this may be found in Ref. [24]. The noise level is taken to be the same for both RC and ML directions, that is $\alpha=\beta=30$.

The spatial 2D distribution of the axons corresponding to labeled RGCs is shown at Figure 7. The "labeling spot" in retina is a circle with radius R=7.3, the coordinates of the center are (15,50), (50,50) and (85,50) on the 100×100 grid. The distribution is obtained by averaging the positions of the labeled axons in SC over 1000 realizations after it reached the stationary solution at $1 \times 10^6$ iterations. The temporal evolution of the map for the label in the central retina is shown at the Figure 8. It corresponds to averaging over 1000 different realizations at each time interval.

In both 1D and 2D cases the calculations were performed on Dell PowerEdge 1600SC server. The programs, written on Matlab (MathWorks, Inc.), are available for download in [23].